# Pressure-Induced Insulating State in $Ba_{1-x}RE_xIrO_3$ (RE = Gd, Eu) Single Crystals

O. B. Korneta[1,2], S. Chikara[1,2], S. Parkin[1,3] L.E. DeLong[1,2], P. Schlottmann[4] and

G. Cao[1,2*]

[1]Center for Advanced Materials, University of Kentucky, Lexington, KY 40506

[2]Department of Physics and Astronomy, University of Kentucky, Lexington, KY 40506

[3]Department of Chemistry, University of Kentucky, Lexington, KY 40506

[4]Department of Physics, Florida State University, Tallahassee, FL 32306

$BaIrO_3$ is a novel insulator with coexistent weak ferromagnetism, charge and spin density wave. Dilute RE doping (e.g., ~ 4%) for Ba induces a metallic state, whereas application of modest pressure ($\leq$ 12.1 kbar) readily restores an insulating state characterized by a three-order-of-magnitude increase of resistivity. Since pressure generally increases orbital overlap and broadens energy bands, a pressure-induced insulating state is not commonplace. The profoundly dissimilar responses of the ground state to light doping and low hydrostatic pressures signal an unusual, delicate interplay between structural and electronic degrees of freedom in $BaIrO_3$.

PACS: 71.30.+h, 75.50.Dd, 75.30. Fv, 62.50.-p

**I. Introduction**

Oxides of the 5d-element Ir should be more metallic and less magnetic than their 3d-element counterparts, owing to the reduced Coulomb interaction U and broadened 5d bandwidth W as a consequence of the relatively more extended 5d orbitals as compared to 3d wave functions. However, radical departures from these intuitive expectations are commonplace among iridates such as $BaIrO_3$ **[1-9]**, $Sr_{n+1}Ir_nO_{3n+1}$ (n = 1, 2) **[10-17]**, $Ca_5Ir_3O_{12}$ and $Ca_4IrO_6$ **[18]**. These magnetic insulators exhibit strong competition between exotic cooperative states, including anomalous "diamagnetism" in $Sr_3Ir_2O_7$ **[15]** and giant magnetoelectricity within a novel $J_{eff}$ = 1/2 Mott state **[16,17]** in $Sr_2IrO_4$ **[19]**. These novel phenomena are manifestations of the strong spin-orbit coupling and d-p hybridization that vigorously compete with other interactions. Since structural distortions can lift orbital degeneracy to effectively narrow the 5d-bandwidth **[16,17,19]**, an iridate can be delicately poised near a metal-insulator (MI) boundary, requiring only a slight lattice strain to induce a MI transition.

$BaIrO_3$ stands out because it is extremely sensitive to lattice contraction **[5]** and has been found to exhibit a simultaneous onset of weak ferromagnetism (WFM) and charge density wave (CDW) order at a Curie temperature $T_C$ (=183 K) **[3]**. This phase has been confirmed by other independent studies **[4, 6, 7, 9]**. A space modulation of the charges in a ferromagnet automatically generates also a spin density wave (SDW), so that this complex state should have three order parameters for the CDW, the SDW and the magnetization. This complex phase transition is unique to $BaIrO_3$. We expect the exotic WFM/CDW/SDW state to behave extraordinarily when probed with small external perturbations.

In this paper, we report the effects of dilute rare earth doping and modest hydrostatic pressure on the structural, thermal, magnetic and transport properties of single-crystal $Ba_{1-x}RE_xIrO_3$ (RE = Gd, Eu).  The RE impurities introduce (i) donor states and (ii) induce lattice strains, which both strongly affect the electronic properties.  The most salient features of the compound are:  **(1)** A ***metallic state*** with intriguing magnetic and transport behavior is rapidly induced with dilute rare earth doping (e.g. ~ 4% Gd for Ba) of $BaIrO_3$, which has a soft gap of ~ 0.15 eV **[3,7]**.  **(2)** The metallic state is readily driven back into an ***insulating state*** with application of modest pressures (P ≤ 12.1 kbar) that prompts an unusually rapid increase in resistivity ρ by as much as ***three orders of magnitude***.  A pressure-induced insulating state has only been observed in a handful of materials, and ***only under extremely high pressure*** (50 - 950 kbar); examples include elemental Li **[20-21]** and $La_{1-x}Sr_xCoO_3$ **[22]**.  Such a phenomenon has not been observed in a 5d-based material in any range of pressure, and is shown herein to be a consequence of unusual electron-lattice interactions in $BaIrO_3$.

As a consequence of the CDW order $BaIrO_3$ is an insulator and impurities give rise to bound states inside the gap.  In particular, trivalent rare earth ions substituting for divalent Ba ions yield donor states. With increasing concentration the donor bound states will overlap and eventually form an impurity band.  For concentrations x larger than an $x_c$ the system will be metallic, i.e. there is an insulator-metal transition at $x_c$. $Gd^{3+}$ having an ionic radius smaller than $Ba^{2+}$ induces as well a "negative pressure" on the system, distorting this way the Ir-O bond angles.  The extreme instability of the electronic state of $BaIrO_3$ is revealed by two diametrically opposed responses to lattice strain.  The system first becomes metallic under doping and the insulating state is restored with hydrostatic pressure.

**II. Experimental**

Single crystals were grown using flux techniques described elsewhere [3, 5]. The average size of the single crystals was approximately $0.5 \times 0.5 \times 0.8$ mm$^3$ with the longest dimension along the **c**-axis. The crystal structure was determined at both 295 K and 90 K from a small fragment ($0.05 \times 0.05 \times 0.05$ mm$^3$) using Mo K$\alpha$ radiation and a Nonius Kappa CCD single-crystal diffractometer. All doping concentrations and chemical compositions of single crystals studied were determined by Energy Dispersion x-ray (EDX). Measurements of heat capacity C, magnetization M and resistivity $\rho$(T < 400 K) were performed using either a Quantum Design PPMS or MPMS SQUID Magnetometer equipped with a Linear Research Model 700 AC bridge and a pair of Keithley 2182 and 2400 for transport measurements. The thermoelectric power S was measured using a homemade apparatus. High-temperature S(T) and $\rho$(T) were measured using a closed-cycle displex cryostat operating from 9 K to 900 K. Hydrostatic pressure measurements of $\rho$(T) to 13 kbar were conducted using a piston-in-cylinder Cu-Be clamp pressure cell and low-viscosity "Daphney 7373" oil as a pressure medium.

**III. Results**

The *C2/m* space group of BaIrO$_3$ features three face-sharing IrO$_6$ octahedra forming Ir$_3$O$_{12}$ trimers that are vertex-linked via IrO$_6$ octahedra (containing *Ir1* and *Ir3* sites) to constitute one-dimensional (1D) chains along the **c**-axis. A monoclinic distortion generates twisting and buckling of the trimers (which are tilted ~12º relative to each other) [1, 2] that gives rise to two 1D zigzag chains along the **c**-axis, and a 2D layer of corner-sharing IrO$_6$ octahedra in the ab-plane (see **Fig.1**) **[1-3, 7]**. Dilute trivalent Gd$^{3+}$ (ionic radius $r_{Gd}$ = 0.94 Å) substitutions for divalent Ba$^{2+}$ ($r_{Ba}$ = 1.35 Å) preserve the crystal structure, but significantly change the temperature-dependent lattice parameters (see **Table 1**) that prove critical to the novel behavior reported here.

**Table 1** *Lattice parameters and Ir1-O2-Ir3 bond length and angle at 295 K (90 K)*

| 295 K (90 K) | a [Å] | b [Å] | c [Å] | β | Ir1-O [Å] | Ir3-O [Å] | Ir1-O-Ir3 (°) |
|---|---|---|---|---|---|---|---|
| x=0.00 | 10.0180 (9.9907) | 5.7515 (5.7342) | 15.1867 (15.2359) | 103.2920° (103.3951°) | 1.9930 (1.9910) | 2.0250 (2.0240) | 163.0° (161.6°) |
| x=0.04 | 9.9978 (9.9852) | 5.7386 (5.7381) | 15.0021 (15.2028) | 103.2200° (103.3514°) | 1.9760 (1.9880) | 2.0170 (2.0230) | 163.5° (162.3°) |

Gd doping induces pronounced changes in a wide range of physical properties of single-crystal Ba$_{1-x}$Gd$_x$IrO$_3$ ($0 \leq x \leq 0.06$), as documented by representative data for M(T), C(T), ρ(T) and S(T) shown in **Fig. 2**. T$_C$ is suppressed from 183 K for x = 0, to 140 K for x = 0.06 (**Fig. 2a**). The Curie-Weiss temperature θ$_{cw}$ (extrapolated from $\Delta\chi^{-1}$ = $[\chi(T) - \chi_o]^{-1}$, where $\chi_o$ is T-independent), obtained from data in the range 200 < T < 350 K, changes sign with doping: θ$_{cw}$ = +115 K (-52 K) for x = 0 (0.04), signaling a change in exchange coupling from FM for x = 0 to antiferromagnetic (AFM) for x = 0.04 (**Fig. 2a**). In addition, a new, sharply defined phase transition occurs at T$_M$ = 9.5 K in x = 0.04, and persists (although significantly weakened) for x = 0.06, as discussed below.

All the phase transitions evident in the magnetic data are reflected in C(T), as shown in **Fig. 2b**. A small specific heat anomaly ($|\Delta C| \sim$ 2 J/mole K) is observed near T$_C$ for x = 0, but becomes barely noticeable for x = 0.04. C(T) for x = 0.04 exhibits rather unusual and complex behavior below $T_M$, as discussed below.

Gd-doping also strongly impacts the transport properties, as indicated in **Fig. 2c**. The temperature dependence of the resistivity in the ab-plane, $\rho_{ab}(T)$, exhibits a sharp kink at $T_C$ = 183 K for x = 0, consistent with early results [3]. This transition has been associated with the formation of a CDW gap ($E_g$ ~ 0.15 eV and 0.13 eV for current along the ab-plane and the c-axis, respectively), so that BaIrO$_3$ is in an insulating state below $T_C$. The gap can be readily reduced by merely a few percent of RE doping (the effect of Gd and other RE doping on $\rho$ is similar). This leads to an insulator-metal transition as a function of x as seen in **Fig. 2c**. Data for $\rho_{ab}(T)$ for x = 0.04, 0.06 and 0.07 show both $T_C$ and the gap energy systematically decrease with x. At high temperatures the resistivity first increases with x and then it decreases rapidly again; this behavior is consistent with an alloy in which one component (Gd$^{3+}$) contributes with one more electron than the other (Ba$^{2+}$).

The resistivity of Ba$_{1-x}$Gd$_x$IrO$_3$ is consistent with the ab-plane thermopower $S_{ab}(T)$, which exhibits a sharp transition at $T_C$ = 183 K for x = 0 that is followed by a rapid, 5-fold increase that peaks near 75 K, confirming the gap opening at $T_C$. The magnitude of $S_{ab}$ for x = 0 is comparable to those of other insulating iridates **[19]**. $S_{ab}(T)$ for x = 0.04 exhibits a similar temperature dependence with a broadened transition at $T_C$ = 154 K, and a smaller overall magnitude (the peak value $S_{ab}$ ~70 µV/K for x = 0.04 is only 1/3 of that (250 µV/K) for x = 0), as shown in **Fig. 2d**. A reduced $S_{ab}(T)$ is consistent with a robust metallic state, since $S \propto \rho_F$ ($\rho_F$ is the density of states) at low T [23-26]. Above $T_C$, $S_{ab}(T) \propto 1/T$ over a range $190 < T < 310$ K for $x = 0.04$. The positive sign of $S_{ab}(T)$ indicates that charge carriers are primarily holes.

The above low T results are roughly consistent with trivalent $Gd^{3+}$ donor states as known for ordinary semiconductors. The concentrations discussed here are beyond the insulator-metal transition of the impurity band and hence the doped system should behave like a bad metal. The change of the ab-plane thermopower is negative from x=0 to 0.04, as expected from the electron donor states of trivalent $Gd^{3+}$.

In **Fig. 3** we address the low temperature behavior of the system. The low-T metallic state (x=0.04) exhibits complex thermodynamic behavior, as seen in **Fig. 3a**, where a phase transition is evident at $T_M = 9.5$ K in $M_{ab}(T)$; this anomaly is followed by two more transitions at $T_{M2} = 6.5$ K and $T_{M3} = 3$ K, respectively. Three transitions are also clearly reflected in C(T), confirming a bulk effect. A fourth anomaly is seen in C(T) near 11.5 K, but not in M(T). Subtraction of C(T<15K) of pure $BaIrO_3$ as a background yields $\Delta C_M(T)$, which is attributed to the magnetic 4f-orbital contribution, and is shown in **Fig. 3b**. The magnetic entropy removal $S_M \approx 0.95$ J/mole-K is calculated from $\Delta C_M(T)$ (assuming $S_M = R\ln(2S+1)$, $R \equiv$ gas constant and $S = 7/2$) and corresponds to approximately 5% $Gd^{3+}$ ions, in reasonable agreement with the sample composition $x = 0.04$ determined by EDX. This indicates that all magnetic transitions below $T_M$ are due chiefly to Gd moments. The Sommerfeld coefficient $\gamma$ is drastically enhanced above $T_M$ from $\gamma = 1$ mJ/mol $K^2$ for $x = 0$ to an extrapolated $\gamma = 50$ mJ/mol $K^2$ for $x = 0.04$ (see dashed line in **Fig. 3c**). We also found that $\gamma = 12$ mJ/mol $K^2$ for 4% Eu doping (not shown). The enhanced $\gamma$ suggests a sizable increase in the density of states at the Fermi energy which can arise either from the Gd donor states, or from metallic 5d-electrons, or from a contribution from magnetic fluctuations that persist for temperatures between the three magnetic phase transitions.

C(T)/T decreases progressively with increasing H (see **Fig. 3c**), and $T_M$, $T_{M2}$ and $T_{M3}$ become less well-defined for $\mu_o H > 3$ T, indicating a gradual polarization of the spin structure that is punctuated by two consecutive metamagnetic transitions observed in the isothermal magnetization $M_c$ at $H_{c1} = 0.2$ T and $H_{c2}$ 1.8 T, and T = 1.7 K (see **Fig. 3d**). The saturation moment extrapolated to H = 0 is ~ 0.3 $\mu_B$/f.u. This value is over 10 times as large as that observed for x = 0, and corresponds to a 4.3 % Gd contribution (i.e., 7 $\mu_B$/Gd$^{3+}$ expected for S = 7/2). Remarkably, we observe that doping other rare earths, such as Pr, Sm, Eu, and Lu, fails to induce the strong magnetic anomalies observed for Gd substitutions. Since these RE elements span a significant range of 4f-electron crystal field schemes, localization and ionic radius, the unique effects of Gd substitution must arise from the maximal de Gennes factor and/or the zero orbital angular momentum contribution (L = 0) to the Gd$^{3+}$ magnetic moment. It is noted that the other RE elements (except Lu) can exhibit anisotropic (orbital) exchanges since L$\neq$0, while Gd cannot.

The transport behavior of Ba$_{0.96}$Gd$_{0.04}$IrO$_3$ is defined by two anomalies in $\rho$, namely a broad peak at T* ≈ 250 K and an abrupt change of slope at $T_C$, as shown in **Fig. 4a**. The temperature dependence of $\rho_{ab}$ bears a striking resemblance to that of the classic CDW material, NbSe$_3$ **[27]**. The broad peak at T* corresponds to no magnetic anomaly on other physical properties and is completely insensitive to applied fields $\mu_o H \leq 7$ T, which suggest that scattering to phonon modes (i.e., electron-lattice coupling) should be considered rather than spin scattering. On the other hand, the transition signature in $\rho(T\approx T_C)$ is sharp and sensitive to H. An abrupt increase of $\rho_{ab}$ immediately below $T_C$ is followed by a gradual decrease, signaling that the Fermi surface is only partially removed by the CDW at $T_C$. The application of a magnetic field $\mu_oH = 7$ T upshifts $T_C$ by 7 K (from 154 K to 161 K), and implies that electronic Zeeman energy is sufficient to alter the gapping of the Fermi surface. This could occur through an improved nesting condition of the Fermi surface or a strengthened electron-phonon coupling.

These results pose an interesting dichotomy: Gd doping lowers the CDW transition and promotes a metallic state, but an applied field raises $T_C$, implying the field increases the gap (the $Gd^{3+}$ moments possibly enhance this effect). In contrast, the energy gap (~ 0.15 eV at $T_C = 183$ K **[3]**) of pure $BaIrO_3$ apparently dominates the external magnetic field, since both the resistivity and magnetization show no visible field dependence near $T_C$ **[3,5]**. It is also noted that a minimum occurs in $\rho_{ab}(T)$ near 20 K but diminishes with x (see **Fig. 2c**), suggesting that it is not due to doping-induced lattice disorder. The insensitivity of the minimum to applied field indicates an electron-lattice coupling rather than a magnetic or spin fluctuation interaction. It cannot be ruled out that it may be related to the anomaly in C(T) at 11. 5 K. In contrast to Gd doping, for 4% Eu doping a sharp downturn rather than an upturn is seen in $\rho_{ab}(T)$, while $\rho_c(T)$ displays the expected downturn as for Gd. From the field-dependence of the magnetization in **Fig.3d** we expect Eu to be trivalent or intermediate valent, i.e. it does not have a free moment as $Eu^{2+}$. The above phenomena further highlight the exotic nature of the ground state in the iridate.

Given the unusual sensitivity of $BaIrO_3$ to lattice contraction, a more metallic state should be established when hydrostatic pressure P is applied to $Ba_{1-x}Gd_xIrO_3$: The essential effect of pressure is to increase the overlap of adjacent electronic orbitals, which can broaden energy bands in a way that supports a metallic state. This is indeed observed in a wide variety of materials---particularly in 3d-transition-metal oxides **[22, 29-31 and references therein]**. In particular, the effect of pressure on the perovskite manganites is primarily to increase the Mn-O-Mn bond angle, but reduce the Mn-O-Mn bond length, thus broadening the band and stabilizing the FM metallic state **[29, 30]**. An exception is Yb compounds **[32-34]**, where pressure reduces the hybridization of the 4f-hole with the ligand states and leads to an increased localization of the 4f-hole.

It is therefore astonishing that application of modest pressures readily drives $Ba_{1-x}Gd_xIrO_3$ and $Ba_{1-x}Eu_xIrO_3$ back into an insulating state with as much as a *three-order-of-magnitude* increase of $\rho_c$ at T = 1.8 K and P ≤ 12.1 kbar, as shown in **Figs. 4b** and **4c**. Our early pressure study **[35]** shows that $\rho_c$ for x = 0 rises by a factor of 30 at T = 1.8 K and P = 12.8 kbar; and $T_C$ (as seen in both $\rho_c$ and M) monotonically decreases at a rate of $dT_C/dP$ = -1.7 K/kbar; a similar but weaker pressure effect on $\rho$ for polycrystalline $BaIrO_3$ was also reported **[36].**

### IV. Discussion

Many features in the CDW insulator $BaIrO_3$ are intriguing and yet to be understood but one major point central to this work is clearly established: a robust metallic state is induced by merely 4% Gd doping of $BaIrO_3$ (the effect of all other RE ion doping, such as Pr, Eu, Sm and Lu, on the *transport properties* is essentially the same). A pressure of 10 to 12 kbar reinstates the insulating ground state.

Although $BaIrO_3$ is not a simple semiconductor, but the insulating state at low T is the consequence of a complex ground state with three order parameters (CDW, SDW and the magnetization), the transition to a metallic state via donor doping can be understood in analogy to ordinary semiconductors, i.e. overlapping donor states that eventually form a conducting impurity band. In this case the application of pressure should increase the overlap and lead to better conductivity. This is in contrast to our observations since pressure restores the insulating state. Hence, the physics associated with the doped system is more complex than expected for ordinary semiconductors.

We argue that strong d-p hybridization in the iridate results in unusual electron-lattice effects that are responsible for the anomalous behavior described herein. Band structure calculations for $BaIrO_3$ [4] show that a sharp peak in the density of states near the Fermi energy is governed by the corner-sharing $IrO_6$ octahedra that contain *Ir1* and *Ir3* sites. In addition, there is nesting of the Fermi surface [4], so that many states on the Fermi surface can be simultaneously affected by a lattice distortion of a given wavevector. It is therefore reasonable that isostructural 9R-$BaRuO_3$ [1, 2], which suffers no monoclinic distortion or octahedral buckling that exists in $BaIrO_3$, has a robust metallic state exhibiting strong quantum oscillations [28]. This view is also corroborated by the behavior of single-crystal $Ba_{1-x}Sr_xIrO_3$, where Sr ($r_{Sr}$ = 1.18 Å) substituting for Ba ($r_{Ba}$ = 1.35 Å) relaxes the twisting and buckling of the $IrO_6$ octahedra. Here $T_C$ is drastically suppressed and eventually the metallic state is recovered at x = 0.12 [5]. Gd ($r_{Gd}$=0.94 Å) (or other RE) doping for Ba apparently has a profound effect on the Ir – O bond angles. $BaIrO_3$ is a rare example of a material that is extremely close to a metal-insulator borderline and susceptible to lattice contraction.

The pressure-induced insulating state we observe in $Ba_{1-x}Gd_xIrO_3$ is a fundamentally new phenomenon, as it happens in a material with extended 5d-orbitals and at relatively low pressures. Two recent examples of a pressure-induced insulating/semiconducting state are Li at P = 950 kbar (after undergoing several phase transitions to low-symmetry structures) **[21]**; and (La,Sr)CoO$_3$, which becomes an insulator for P ≈ 140 kbar due to a change in the spin state of the Co$^{3+}$ ion **[22]**. In both cases, the necessary pressures are much higher than those applied in this work (P ≤ 12.1 kbar).

A careful examination of crystal structure data (**Table 1**) reveals important changes in lattice stability at lower temperatures due to 4% Gd doping: **(1)** The *Ir1-O2-Ir3* bond length (*Ir1-O2* and *Ir3-O2*) that links the corner-sharing IrO$_6$ octahedra grows considerably with decreasing T for x = 0.04, but shrinks slightly for x = 0. **(2)** The *Ir1-O2-Ir3* bond angle that reflects the tilting and twisting of the corner-sharing IrO$_6$ octahedra is significantly larger for x = 0.04 (162.3°) than for x = 0 (161.6°) at 90 K (see **Fig. 5a**). **(3)** While the unit cell volume shrinks as T is lowered from 295 K to 90 K, the c-axis grows by more than 1.34% for x = 0.04, but only 0.30% for x = 0.

All the above changes illustrate that the lattice stability is significantly altered, and that the buckling of corner-sharing Ir$_3$O$_{12}$ trimers is considerably weakened (particularly at lower temperatures), with only dilute RE doping. Since the t$_{2g}$-block bands sensitively depend on *Ir1 and Ir3* atoms and their d-orbitals **[4]**, any reduction of the twisting and tilting of corner-sharing IrO$_6$ octahedra that contain *Ir1* and *Ir3* inevitably leads to the enhancement of the density of states near the Fermi surface (evident in the Sommerfeld γ coefficient). Together with the donor impurity band it drives the metallic state below T$_C$. **Fig. 5a** schematically illustrates the change of the *Ir1-O2-Ir3* bond angle with x at T = 90 K and its critical correspondence to the electronic structure of the iridate: *Ir1-O2-Ir3 bond angles closer to ideal value of 180° favor a* *metallic state*. Moreover, a relaxed *Ir1-O2-Ir3* bond angle is intimately associated with a sizable elongation of the c-axis as T is lowered (negative expansion), which is accompanied by a positive thermal expansion of the a- and b-axes, and signals a lattice with weakened stability.

The above lattice anomalies suggest that **under pressure** the three $t_{2g}$ orbitals are affected **significantly differently**. Consequently, the effects of **hydrostatic pressure** on the $t_{2g}$ orbitals may appear to be paradoxically **uniaxial or anisotropic**. It is therefore reasonable to expect that with increasing P, the *Ir1-O2-Ir3* bond angle is severely bent, causing the electronic structure to revert to an insulating state that is even stronger than that in the x = 0 composition. **Fig. 5a** illustrates the intimate correlation of the *Ir1-O2-Ir3* bond angle with $\rho_c$, which rises by three orders of magnitude as P increases to 12.1 kbar for $Ba_{0.96}Gd_{0.04}IrO_3$ at T = 1.8 K.

### V. Conclusions

A robust metallic state is induced by merely 4% Gd doping of $BaIrO_3$ (the effect of all other RE ion doping on the resistivity is essentially the same), while a pressure of 10 to 12 kbar reinstates the insulating ground state. The former can be understood with an impurity band formation in a semiconductor, but the pressure dependence contradicts this picture. It is necessary to invoke changes in bond lengths and angles to explain a pressure induced metal-insulator transition.

The doping and pressure effects on the electronic ground state are summarized in **Figs. 5b** and **5c**. The novelty of this system lies in the observation that the *weak* lattice contraction driven by RE doping and pressure prompts **profound** and yet entirely **opposite** changes in the electronic state of the iridate. The extraordinary delicacy of the ground state underlines a new, critical balance between various orbital, electronic and lattice degrees of freedom.

**Acknowledgements:** G.C. is pleased to acknowledge very useful discussions with Dr. Mike Whangbo. This work was supported through NSF grants DMR-0552267, DMR 0856234 and EPS-0814194. LED's research is supported by DoE Grant #DE-FG02-97ER45653. PS is supported by the DoE through Grant #DE-FG02-98ER45707.

**\***Corresponding email: cao@uky.edu

**Captions**

**Fig.1.** Crystal structure highlighting $Ir_3O_{12}$ cluster trimers along the c-axis.

**Fig.2.** The temperature dependence up to 600 K of **(a)** the magnetization M(T) and $1/\Delta\chi$ (right scale, data points and Curie-Weiss fit for x=0), **(b)** specific heat C(T), **(c)** $\rho_{ab}$(T) (right scale for x=0.06 and 0.07) and **(d)** $S_{ab}$(T) for $Ba_{1-x}Gd_xIrO_3$ with $0 \leq x \leq 0.07$. Inset in **Fig.2b**: enlarged C(T) near $T_C$. The dashed line extrapolates the background.

**Fig.3.** The low temperature dependence for $Ba_{1-x}Gd_xIrO_3$ of **(a)** C(T)/T for x=0 and 0.04 and $M_{ab}$(T) for x=0.04 (right scale), **(b)** magnetic contribution to specific heat $\Delta C_M$(T)/T and magnetic entropy removal $S_M$ (right scale) for x=0.04, **(c)** C(T)/T vs. $T^2$ for x=0 and 0.04 at magnetic fields H up to 9 T, and **(d)** the c-axis isothermal magnetization $M_c$ at T=1.7 K for x=0.04 and other rare earth ion doping as indicated (right scale).

**Fig.4.** **(a)** The ab-plane and the c-axis resistivity, $\rho_{ab}$(T) and $\rho_c$ (T), for at $\mu_oH$=0 and 7 T and the ab-plane and c-axis magnetization $M_{ab}$ (T) and $M_c$ (T) at $\mu_oH$=0.01 T for x=0.04 (right scale); **(b)** $\rho_c$(T) for x=0.04 at hydrostatic pressure P up to 12.1 kbar, and **(c)** $\rho_c$(T) for x=0.04 (Eu) ($Ba_{1-x}Eu_xIrO_3$) at P up to 11.8 kbar. Inset in **Fig.4b**: enlarged $\rho_c$ (T) for x=0.04 highlighting the changes near $T_C$. Inset in **Fig.4c**: $\rho_c$(T) and $M_c$ (T) (right scale) for x=0.04 (Eu).

**Fig.5.** **(a) Left Panel**: the schematic of the Ir1-O2-Ir3 bond angle vs. concentration x at T=90 K (the value of the angle for x=0 and x=0.04 are marked on the vertical axis). Note that the Ir1-O2-Ir3 bond vertex-links the trimers; **Right Panel**: $\rho_c$ for x=0.04 vs. P at T=1.8 K and 20 K. Inset: the schematic of the Ir1-O2-Ir3 bond angle in dense x=0.04; **(b)** The T-x phase diagram and **(c)** the T-P phase diagram for $Ba_{1-x}Gd_xIrO_3$. Note: PM stands

for paramagnet, PM-M paramagnetic metal, PM-I paramagnetic insulator, M-M magnetic metal, and M-I magnetic insulator.

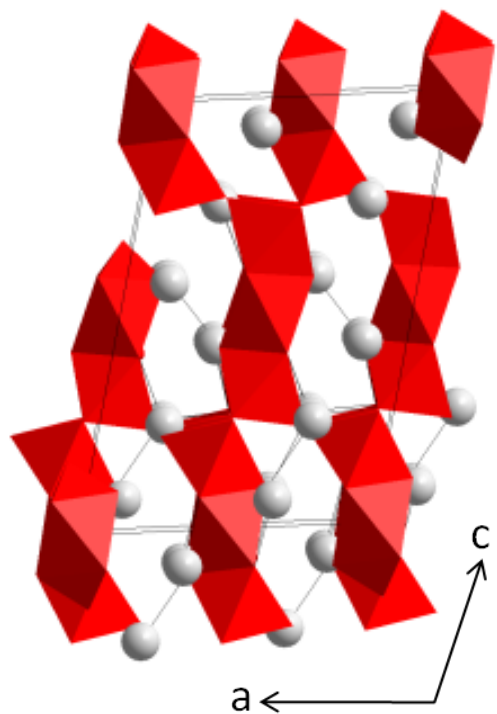

Fig. 1

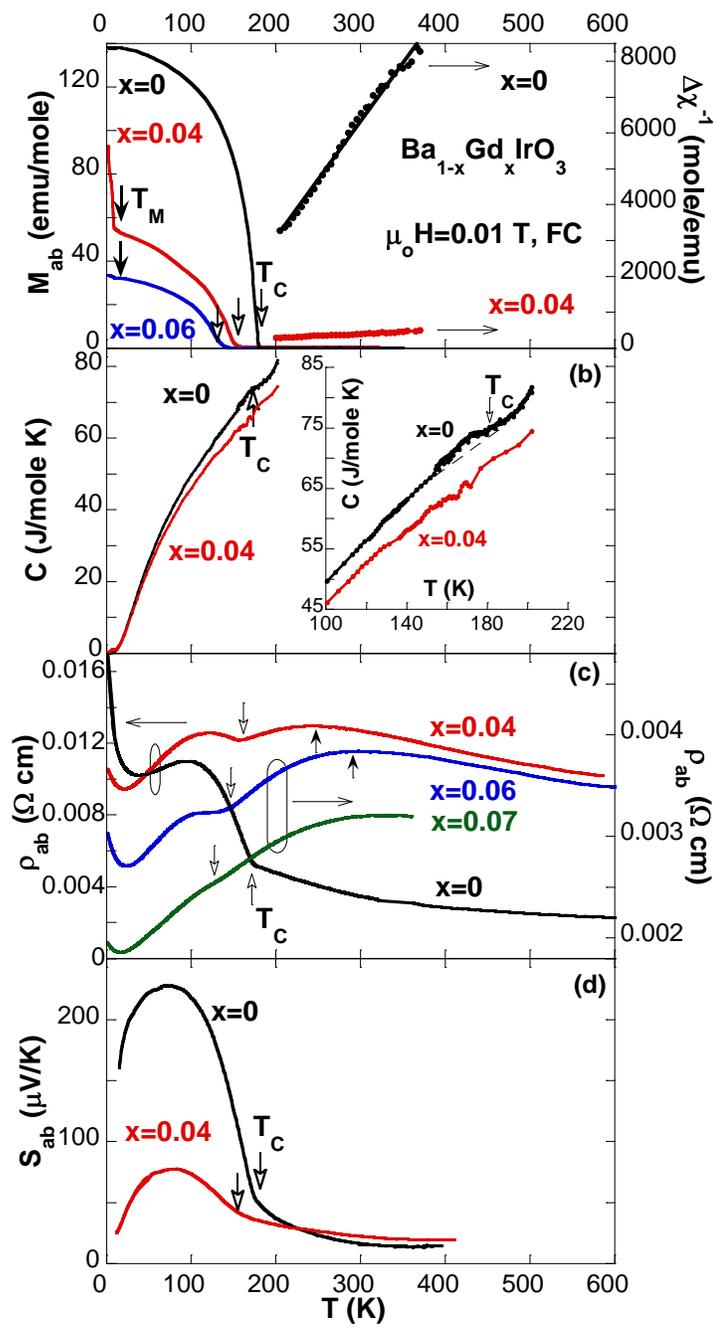

Fig.2

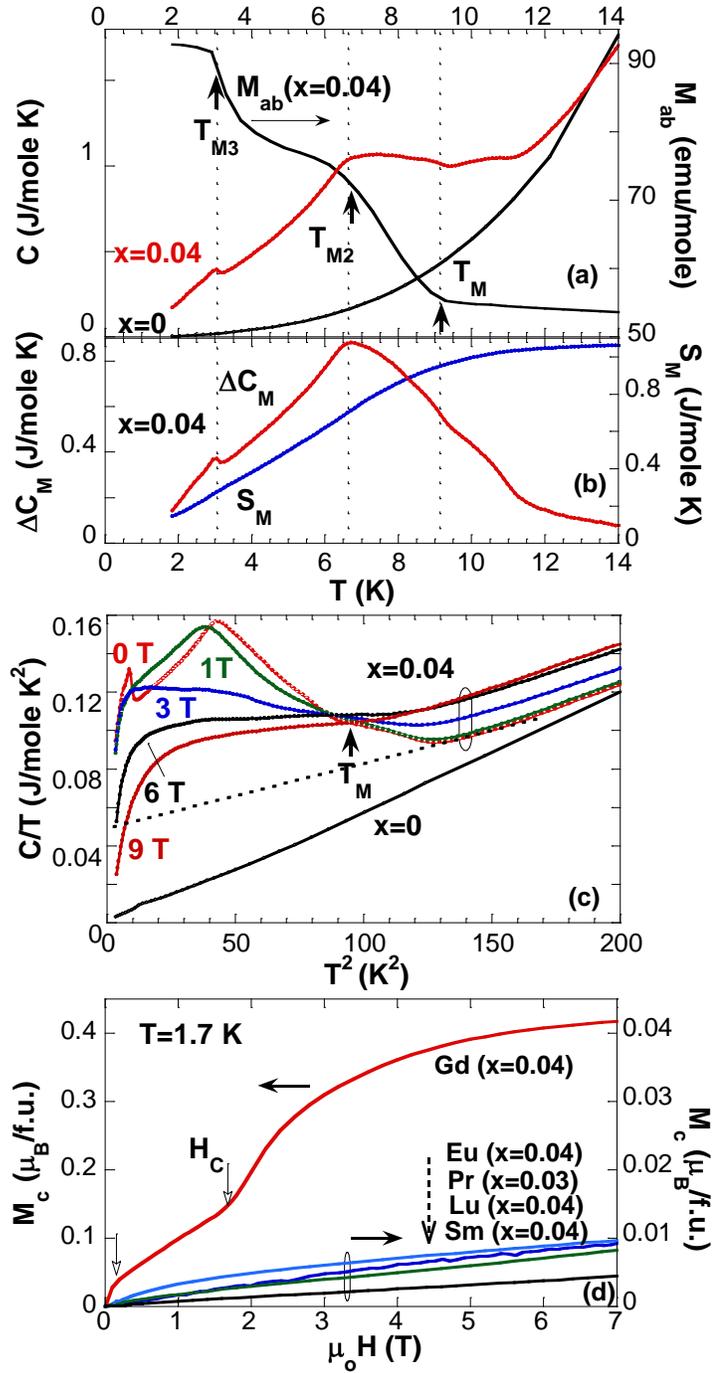

Fig.3

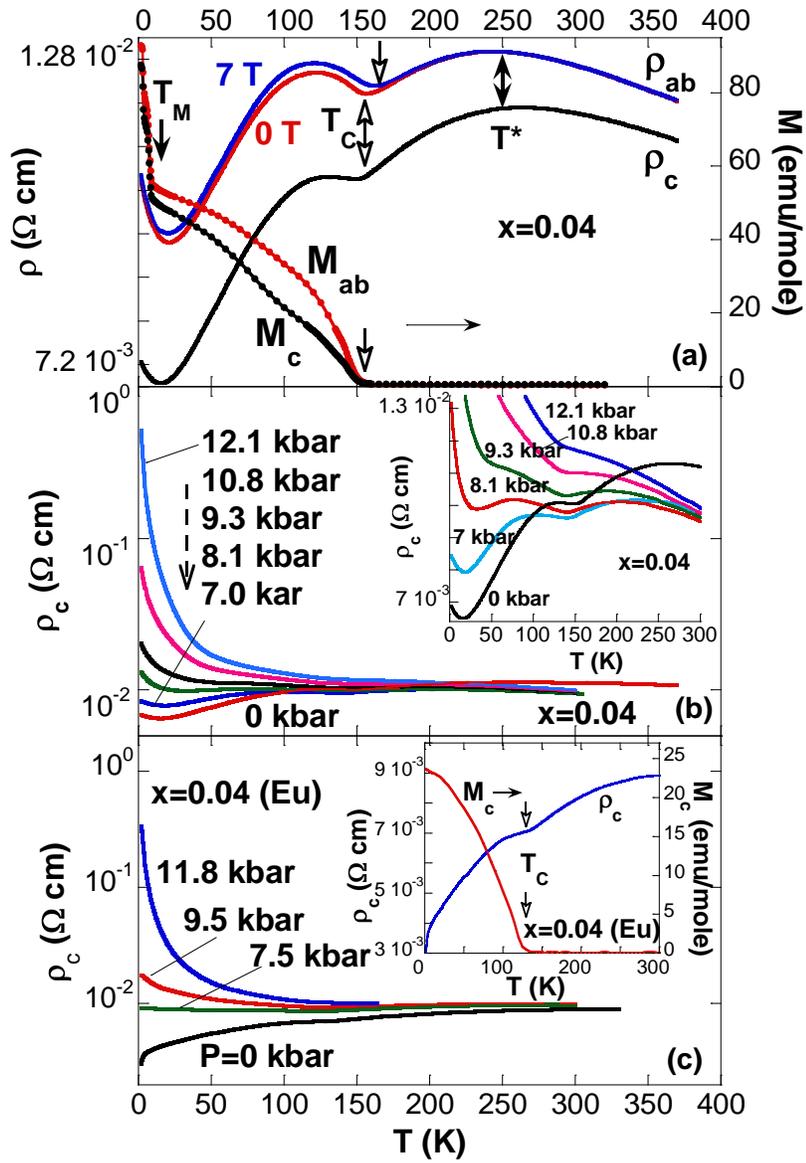

Fig.4

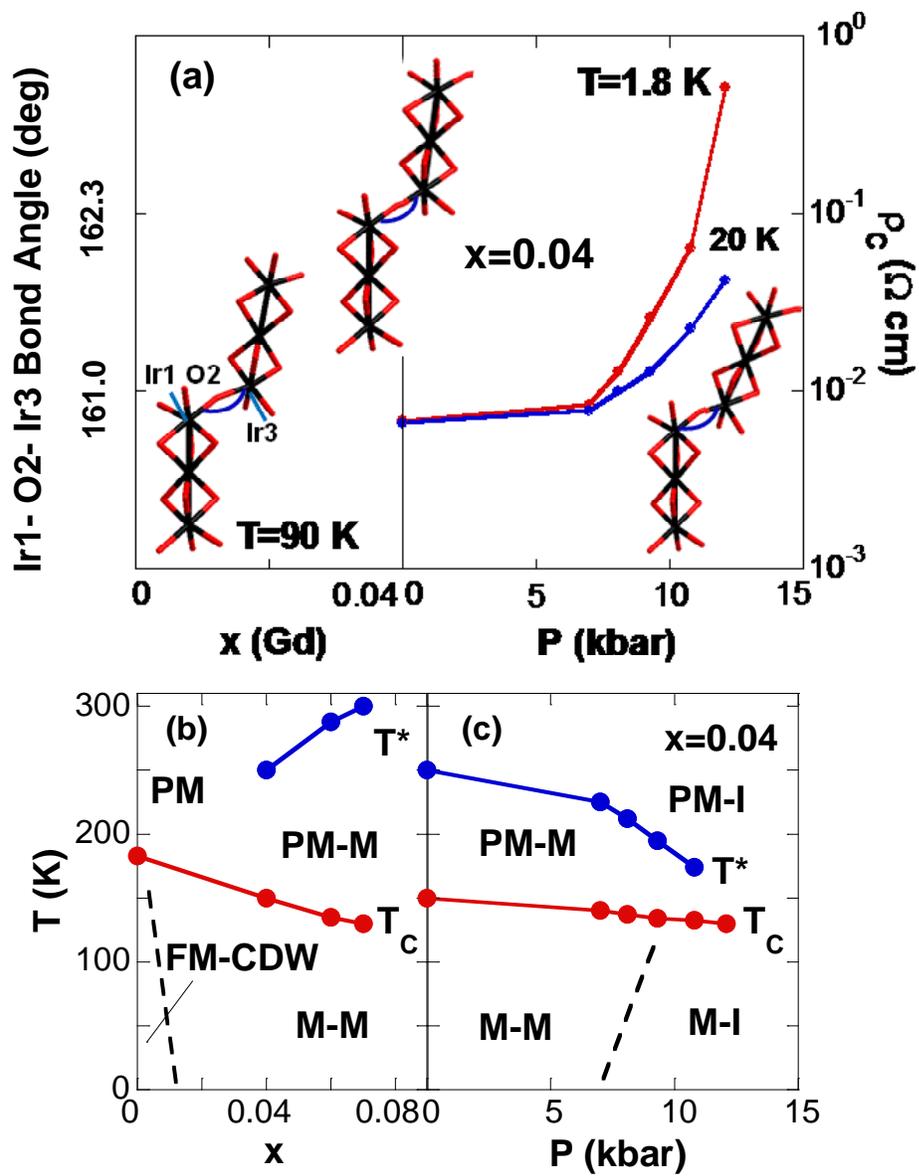

Fig.5